\documentclass[3p,twocolumn]{elsarticle}

\usepackage{amssymb}
\usepackage{booktabs}
 \journal{arXiv}

\usepackage[utopia]{mathdesign}
\usepackage{textcomp}
\usepackage{comment}
\usepackage{subfig,pgfplots,epsf,graphicx,color}
\usepackage{amsmath,amsfonts,amsthm,multirow}

\usepackage{rotating}
\usepackage{tabularx}
\usepackage{float}

\usepackage{units}
\usepackage{pdfpages}
\usepackage{afterpage}
\usepackage{wrapfig}
\usepackage{microtype}
\usepackage{fullpage}
\usepackage{bm} 
\usepackage{caption} 
\usepackage[colorlinks=true]{hyperref}
\hypersetup{
  colorlinks  = true, %Colours links instead of ugly boxes
  urlcolor    = blue, %Colour for external hyperlinks
  linkcolor   = blue, %Colour of internal links
  citecolor   = blue, %Colour of citations
}

\DeclareGraphicsExtensions{.gif,.png,.jpg,.pdf,.tiff,.eps}	
 \usepackage{lineno,hyperref}
\modulolinenumbers[5]

\usepackage{tikz}
\usepackage{pgfplots}
\usepackage{pgfmath}
\usepackage[ruled,vlined,linesnumbered]{algorithm2e}
\DeclareTextCommandDefault{\textcopyright}{\textcircled{c}}
\DeclareTextCommandDefault{\textregistered}{\textcircled{%
      \check@mathfonts\fontsize\sf@size\z@\math@fontsfalse\selectfont R}}

\newcommand{\be}{\begin{equation}}
\newcommand{\ee}{\end{equation}}
\newcommand{\bes}{\begin{equation*}}
\newcommand{\ees}{\end{equation*}}
\newcommand{\bse}{\begin{subequations}}
\newcommand{\ese}{\end{subequations}}

\def\dO{\partial \Omega}

\def\ee{{\hat {\underline e}}}

\def\scriptO{{{\it O}\kern -.42em {\it `}\kern + .20em}}
\def\RR{{{\rm l}\kern - .15em {\rm R} }}
\def\PP{{{\rm l}\kern - .15em {\rm P} }}
\def\L2{{{\sf L}^2}}
\def\H1{{{\sf H}^1}}

\def\PN2{{\PP_{N}-\PP_{N-2}}}

\def\complex{{{\rm C} \kern - .53em {\rm l} \kern + .38em}}

\def\a1{{ | \lambda_{\min} |}}

\def\l1{{   \lambda_{\min}  }}

\def\bu0{{\underline {\bf 0}}}

\def\br{{\bf r}}
\def\bu{{\bf u}}

\def\bx{{\bf x}}

\def\Oh{{\hat \Omega}}

\def\u0{{\underline 0}}
\def\n12{{n_{\frac{1}{2}}}}
\def\t12{{t_{\frac 1 2}}}
\def\n12{{n_{0.8}}}
\def\t12{{t_{0.8}}}

\newcommand{\pp}[2]{\frac{\partial #1}{\partial #2} }

\begin{document}
%=======================================================================

%%- - - - - - - - - - - - - - - - - - - - - - - - - - - - - - - - - - - - - -%%
\begin{frontmatter}



\title{Exascale Simulations of Fusion and Fission Systems}

%%- - - - - - - - - - - - - - - - - - - - - - - - - - - - - - - - - - - - - -%%


% Suggested author list by Steven, modify as appropriate
\author[1]{Misun Min}\corref{mycorrespondingauthor}\ead{mmin@mcs.anl.gov}
\author[1,5]{Yu-Hsiang Lan}
\author[1,5,6]{Paul Fischer}
\author[2,7]{Elia Merzari}
\author[2]{Tri Nguyen}
\author[7]{Haomin Yuan}
\author[3]{Patrick Shriwise}
\author[1]{Stefan Kerkemeier}
\author[4]{Andrew Davis}
\author[4]{Aleksandr Dubas}
\author[4]{Rupert Eardly}
\author[4]{Rob Akers}
%\author[5]{Malachi Phillips}
\author[5]{Thilina Rathnayake}
%\author[8]{Noel Chalmers}
\author[9]{Tim Warburton}

% total author up to 12 limit

\cortext[mycorrespondingauthor]{Corresponding author}
\address[1]{Mathematics and Computer Science, Argonne National Laboratory, Lemont, IL 60439}
\address[2]{Department of Nuclear Engineering, Penn State, University Park, PA 16802}
\address[3]{Computational Science, Argonne National Laboratory, Lemont, IL 60439}
\address[4]{United Kingdom Atomic Energy Authority, Culham Science Centre, Abingdon, Oxfordshire, OX14 3DB, United Kingdom}
\address[5]{Department of Computer Science, University of Illinois at Urbana-Champaign, Urbana, IL 61801}
\address[6]{Department of Mechanical Science and Engineering, University of Illinois at Urbana-Champaign, Urbana, IL 61801}
\address[7]{Nuclear Science and Engineering, Argonne National Laboratory, Lemont, IL 60439}
%\address[8]{Data Center GPU and Accelerated Processing, Advanced Micro Devices, Inc., Austin, TX 78735}
\address[9]{Department of Mathematics, Virginia Tech, Blacksburg, VA 24061}
% or include affiliations in footnotes:

%\ead{@llnl.gov}

%%- - - - - - - - - - - - - - - - - - - - - - - - - - - - - - - - - - - - - -%%

%%- - - - - - - - - - - - - - - - - - - - - - - - - - - - - - - - - - - - - -%%

\begin{abstract}
    We discuss pioneering heat and fluid flow simulations of fusion and fission
energy systems with NekRS on exascale computing facilities, including
Frontier and Aurora.
The Argonne-based code, NekRS, is a
highly-performant open-source code for the simulation of incompressible and
low-Mach fluid flow, heat transfer, and combustion with a particular focus on
turbulent flows in complex domains.  It is based on rapidly convergent
high-order spectral element discretizations that feature minimal numerical
dissipation and dispersion. State-of-the-art multilevel preconditioners,
efficient high-order time-splitting methods, and runtime-adaptive communication
strategies are built on a fast OCCA-based kernel library, libParanumal, to
provide scalability and portability across the spectrum of current and future
high-performance computing platforms.    On Frontier, Nek5000/RS has achieved
an unprecedented milestone in breaching over 1 trillion degrees of freedom with
the spectral element methods for the simulation of the CHIMERA fusion
technology testing platform. We also demonstrate for the first time the use of
high-order overset grids at scale.

\end{abstract}

%%- - - - - - - - - - - - - - - - - - - - - - - - - - - - - - - - - - - - - -%%

%%- - - - - - - - - - - - - - - - - - - - - - - - - - - - - - - - - - - - - -%%
%%- - - - - - - - - - - - - - - - - - - - - - - - - - - - - - - - - - - - - -%%

\end{frontmatter}
 
%%- - - - - - - - - - - - - - - - - - - - - - - - - - - - - - - - - - - - - -%%
%% Goals for the paper:
%% - Share our GPU experience and the motivation behind the choices we've made
%% - What software has been developed by CEED that apps can use for GPU acceleration?
%% - Is our approach extendable to other GPUs and other architectures (ARM)
%% - What has been the impact in real applications?
%%- - - - - - - - - - - - - - - - - - - - - - - - - - - - - - - - - - - - - -%%
%% Plan for paper content (14 pages total):
%%
%% 1. Introduction                    1.0 pages
%% 2. goverming equations             3.0 pages
%% 3. gpu implementation              3.0 pages
%%   .partitioning                        pages
%%   .communication                       pages
%%   .kernel performance                  pages
%%   .precondining                        pages
%%   .code                                pages
%% 4. Application performance         5.0 pages
%%   .single gpu, single node             pages
%%   .fullcore, rod, pebbles              pages
%%   .dns5x5, spacer grid                 pages
%% 5. Conclusions                     0.5 pages
%%    References                      2.0 pages
%%- - - - - - - - - - - - - - - - - - - - - - - - - - - - - - - - - - - - - -%%

%%- - - - - - - - - - - - - - - - - - - - - - - - - - - - - - - - - - - - - -%%
\vspace*{-.2in}
%%%%%%%%%%%%%%%%%%%%%%%%%%%%%%%%%%%%%%%%%%%%%%%%%%%%%%%%%%%%%%%%%%%%%%%%%%
\section{Introduction}             
%%%%%%%%%%%%%%%%%%%%%%%%%%%%%%%%%%%%%%%%%%%%%%%%%%%%%%%%%%%%%%%%%%%%%%%%%%

Advanced fission and fusion energy hold promise as a reliable, carbon-free energy source capable of meeting the United States' commitments to addressing climate change. A wave of investment in fission and fusion power within the United States and worldwide indicates an important maturation of academic research projects into the commercial space. Nonetheless, the design, certification, and licensing of novel fission and fusion concepts pose formidable hurdles to successfully deploying new technologies. Because of the high cost of integral-effect nuclear experiments, high-fidelity numerical simulation is poised to play a crucial role in these efforts.

Within the DOE's Exascale Computing Project (ECP), the ExaSMR program developed high-confidence numerical methods such as computational fluid dynamics (CFD) for thermal-fluid heat and momentum transport. The project successfully executed these simulations on exascale systems and generated highly detailed virtual datasets of operational and advanced concept nuclear reactors, which have the potential to impact a wide range of reactor designs. The results of our simulation campaigns and the achieved performance improvements were summarized in a recent article \cite{sc23}.  
\textit{In the present article, we further extend those fission reactor results, including, for the first time, results on Aurora. We also introduce the first exascale simulation of a facility representative of a fusion breeding blanket, which presents engineering challenges similar to fission systems.}

Historically, because of the complexity and range of scales involved in turbulent flow (1 $\mu$m--10 m), simplifications involving  Reynolds averaging have been necessary. Such approaches, while powerful and useful for design, require modeling closures that are expensive or impossible to develop and validate, especially at the appropriate scale, and lead to design compromises in terms of operational margins and economics.

 Under ECP, the NekRS code, a GPU-enabled spectral element solver for heat and fluid flow, was developed and deployed to all major GPU architectures~\cite{min23a}, leading to transformative changes in the realm of what can be simulated with the introduction of restrictive assumptions.  
To understand the transformative leap enabled by NekRS, it is instructive to examine simulations in pebble beds. Before NekRS, the largest pebble bed calculations in the literature
were of the order of $~1000$ pebbles. In 2022 we reported calculations involving the full  Mark-I Flouride Cooled High Temperature Reactor core ($352,625$ pebbles) \cite{sc22}. Recently, these achievements were extended to simulations involving radiation transport for a full small modular reactor core involving actually a considerably larger mesh size \cite{sc23} breaching for the first time the billion spectral element barrier. We also note that, throughout the ECP,  NekRS has been tested and deployed on three generations of GPU-based supercomputers.

Fusion breeding blankets are crucial components in nuclear fusion devices, designed to harness the energy produced during fusion reactions. These blankets surround the core where fusion occurs, serving multiple essential functions: they help manage heat transfer, providing thermal insulation and converting heat into usable energy; they protect the reactor structure from intense neutron radiation; and they breed tritium. Breeding blankets use materials like lithium to absorb neutrons from the fusion process, which transmute into more tritium, completing the fusion energy fuel  cycle. This technology is key to making fusion power a viable and sustainable energy source, but it presents several challenges. Materials are subject to an intense radiation field and subject to radiation damage. Moreover, the heat flux encountered at the first wall and in the divertor are very intense and among the highest possible, creating a challenging environment \cite{breed1, breed2}.

\begin{figure*}[t]
  \centering
   \includegraphics[width=0.75\textwidth]{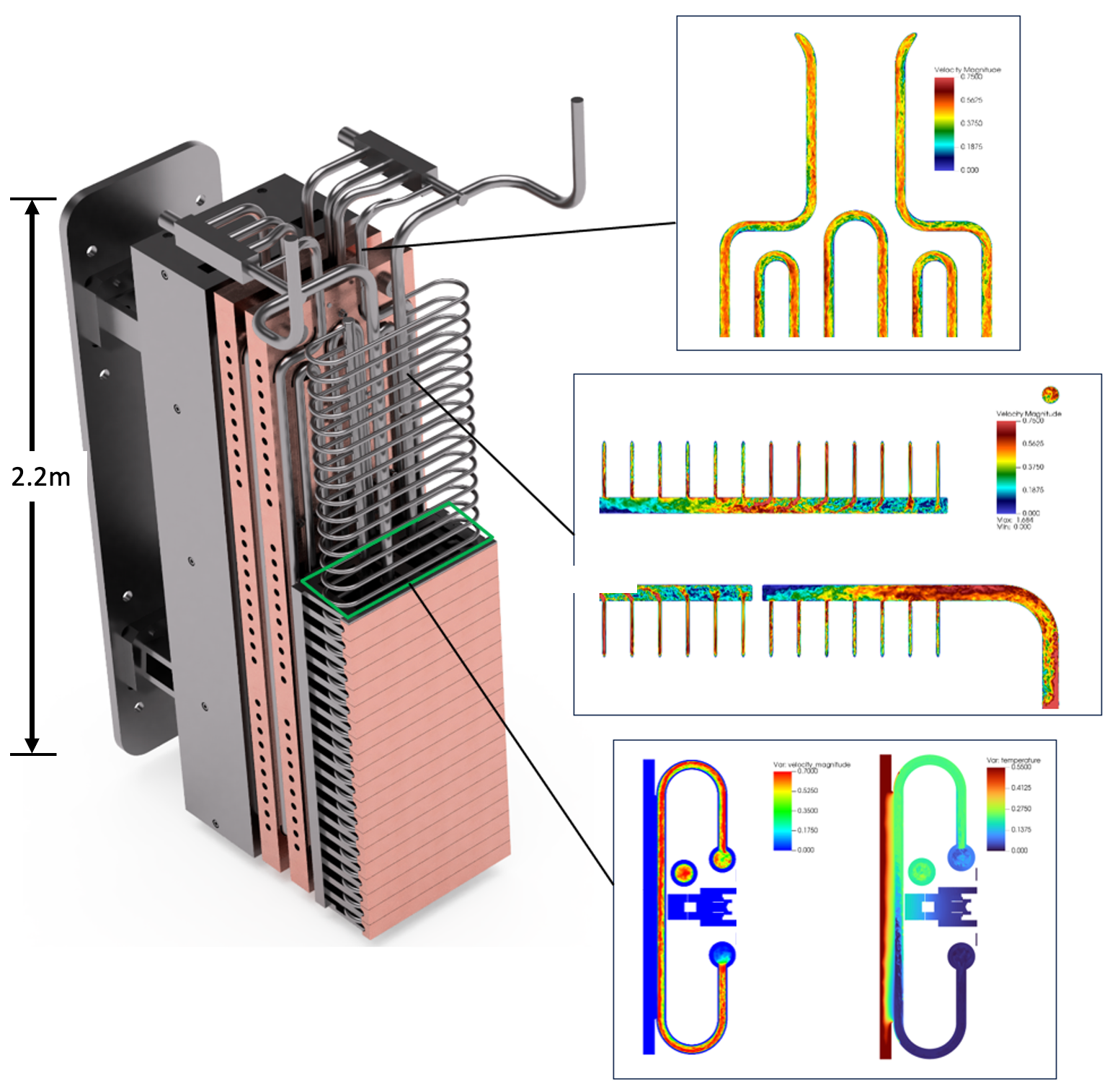}
   \caption{\label{fig:chimera} \small
Rendering of the CHIMERA facility with NekRS simulation results for velocity and temperature. \vspace*{-.1in}}
\end{figure*}

To test various technology candidates, the CHIMERA \cite{chimera} fusion technology facility will enable the testing of large in-vessel component modules under fusion reactor-like conditions of combined in-vacuum thermal power density and magnetic field. With an integral large superconducting magnet and a pressurized water loop, CHIMERA is also ideally placed for experiments on liquid metal breeding blanket prototypes. An infrared heating system has been developed, capable of applying $0.5 MW/m^{2}$ to component surfaces up to the size of the ITER test blanket module first wall. The modules of this heater are highly specialised and designed to endure the high magnetic forces from the CHIMERA static and pulsed magnets. CHIMERA will feature a range of diagnostics, including load cells to measure static and pulsed magnetic forces, and induced current sensors, all of which have been tested to confirm acceptable operation in the pulsed magnetic field. A high-fidelity numerical tool of CHIMERA will support the scientific and engineering goals being pursued.  \textit{We present first-of-a-kind turbulent flow and heat transfer simulations of CHIMERA. Results are shown in Fig.~\ref{fig:chimera}}.

% \begin{figure*}
%   \centering
% \includegraphics[width=0.91\textwidth]{figs/chimera1.png}
%      \caption{\label{fig:chimera} \small
% Rendering of the CHIMERA facility with NekRS simulation results  \vspace*{-.1in}}
% \end{figure*}

CHIMERA includes a complex high Reynolds-number piping system involving several
manifolds and connections. It also involves multiple heating plates and an
infrared system to simulate the intense heat flux of first-wall conditions. The
geometry is exceedingly complex. Generating an adequate discretization
posed significant challenges.
{\em 
To model the system, an overset grid multi-domain approach was used in which
solid and fluid domains are meshed separately, and information is transferred
between the two by interpolation at the boundary \cite{min2018engage, mittal2019}.
   This manuscript represents the first exascale demonstration of
   this technique.
}

To investigate the performance of NekRS on Frontier and Auora, we discuss two numerical models:
\begin{itemize}
\item A portion of a fission core for a Small Modular Reactor. 
\\[-3ex]
\item The CHIMERA model (Fig. \ref{fig:chimera}).
\end{itemize}

\textbf{Fission core.} The core consists of 37 assemblies, each having a
17$\times$17 lattice of fuel pins and guide tubes, with the guide tubes placed in
standard positions. This full-core configuration was modeled in previous publications \cite{fang2021f, sc23} and featured 37 identical assemblies with the same 17$\times$17 configuration. Runs were conducted with up to $6.03 \cdot 10^{11}$  grid points. \textit{In this work, we consider only a subset of the problewm, a single assembly (17$\times$17 configuration), to continue benchmarking the code against novel architectures.} The current fluid mesh for each assembly has $E=27,700$ fluid elements per two-dimensional layer and 
$E=31,680$ solid elements per 2D layer. Multiple extrusions are considered. The fluid
properties used in the simulations were evaluated at $553$ K, and an
incompressible formulation was chosen due to the small changes in density for
the problem of interest. The energy equation is solved in its temperature form.
While the geometry is extruded, each pin has a unique spatially dependent power
profile and composition, making the problem three-dimensional. The work was largely presented in \cite{sc23} for Frontier, but here we present novel results on Aurora. 

\textbf{CHIMERA model.} The mesh for the CHIMERA facility has up to
$E=1,691,729,664$ elements for the piping system and $E=55,056,380$ in the
solid. Several smaller meshes for the piping system were also designed to study
scalability.  The largest mesh is designed for wall-resolved Large Eddy
Simulation and Direct Numerical Simulation of the flow at $Re=10^{5}$--$10^{6}$,
resolving the boundary layer everywhere in the piping. \textbf{This translates
into over $1.1\,\cdot\, 10^{12}$ grid points at polynomial order $N=9$, the largest
simulation ever attempted with the spectral element methods and one of the
largest fluid simulations in a complex geometry to date.}  Inlet-outlet
boundary conditions are applied. In the solid, a heat flux is applied on the
outer surface (i.e., first wall simulator). Heat loads are also applied on some
of  the heat transfer plates, which are illustrated in the 3D relief of
Fig.~\ref{fig:chimera}.

\textbf{We emphasize that the work presented here is a considerable leap in problem size, geometry complexity and simulation complexity (overset grids) compared to our previous submission \cite{sc23}.}

\vspace*{-.08in}
%%%%%%%%%%%%%%%%%%%%%%%%%%%%%%%%%%%%%%%%%%%%%%%%%%%%%%%%%%%%%%%%%%%%%%%%%%
\section{Formulations}            
%%%%%%%%%%%%%%%%%%%%%%%%%%%%%%%%%%%%%%%%%%%%%%%%%%%%%%%%%%%%%%%%%%%%%%%%%%

Heat and Mass transfer are the key mechanisms for power generation in fusion and fission systems.  The problem is governed by
the incompressible Navier-Stokes equations (NSE)
for velocity ($\bu$) and pressure ($p$) coupled with the temperature ($T$) equation,
\small
\begin{eqnarray} \label{eq:nse}
\hspace*{-.30in}
\frac{D T}{Dt}\! :=\! \pp{T}{t} + \bu \cdot \nabla T \!=\!
\frac{1}{Re \cdot Pr} \nabla^2 T  + q \hspace*{.563in} \\[.9ex]
\hspace*{-.30in}
\frac{D\bu}{Dt}\! :=\! \pp{\bu}{t} + \bu \cdot \nabla \bu \!=\!
\frac{1}{Re} \nabla^2 \bu - \nabla p, \;\;\; \hspace*{.200in}
\nabla \cdot \bu = 0,
\end{eqnarray} \normalsize
where $Re=UL/\nu \gg 1$ is the Reynolds number based on flow speed $U$, length
scale $L$, and viscosity $\nu$, and $Pr$ is the Prandtl number (i.e., the ratio
of momentum to thermal diffusivities) and $q$ is the volumetric heat deposition
coming from radiation transport calculations. From a computational
standpoint, the long-range coupling of the incompressibility constraint,
$\nabla \cdot \bu = 0$, makes the pressure substep intrinsically communication
intensive and a major focus of our effort as it can easily consume 60-80\% of
the run time \cite{sc22}. We note that in solids, only the temperature
equation is solved and without the advection term.

NekRS is based on the spectral element method (SEM) \cite{pat84}, in which
functions are represented as $N$th-order polynomials on each of $E$ elements,
for a total mesh resolution of $n=EN^3$.
It originates from two code suites:
  Nek5000 \cite{nek5000}, which is a 1999 Gordon Bell winner \cite{tufo99a},
and
  libParanumal \cite{warburton2019,warburton2019b}, which is a fast
  GPU-oriented library for high-order methods written in the Open Concurrent
  Computing Abstraction (OCCA) for cross-platform portability \cite{occa}.
OCCA provides backends for CUDA, HIP, DPC++, and OpenCL with virtually
no performance degradation over native implementations.
  
Direct numerical simulation (DNS) and large-eddy simulation (LES) of
turbulence typically require simulations over long times to gather statistics.
Campaigns that can last weeks require not only performant implementations but
also {\em efficient discretizations} that deliver high accuracy at low cost per
gridpoint.   Kreiss and Oliger noted early on the importance of high-order
methods when considering long integration times \cite{kreiss72}.  To offset
cumulative dispersion errors, $e(t)\sim C t$, one must have $C \ll 1$ when $t
\gg 1$.    \textbf{For this reason, SEM offers several advantages for the high-fidelity modeling of turbulent heat and fluid flow.} 

SEM accommodates
body-fitted coordinates through isoparametric mappings of the reference
element, $\Oh:=[-1,1]^3$, to individual (curvilinear-brick) elements
$\Omega^e$, $e=1,\dots,E$.  On $\Oh$, solutions are represented in terms of
$N$th-order tensor-product polynomials,
\small \begin{eqnarray} \label{eq:field1}
\hspace*{-.1in}
\left. \bu(\bx) \right|^{}_{\Omega^e} = \sum_{i=0}^N
\sum_{j=0}^N \sum_{k=0}^N \bu_{ijk}^{e}\,h_i(r)\,h_j(s)\,h_k(t),
\\[-3ex] \nonumber
\end{eqnarray} \normalsize
where the $h_i$s are stable nodal interpolants based on the
Gauss-Lobatto-Legendre quadrature points $(\xi_i,\xi_j,\xi_k)\in \Oh$ and
$\bx=\bx^e(r,s,t)$ maps to $\Omega^e$.  This form allows all operator
evaluations to be expressed as {\em fast tensor contractions}, which can be
implemented as BLAS3 operations \cite{sc22} in only $O(N^4)$ work and $O(N^3)$
memory references \cite{dfm02,sao80}.  This low complexity is in sharp contrast
to the $O(N^6)$ work and storage complexity of the traditional $p$-type FEM.
Moreover, hexahedral (hex) element function evaluation is about six times
faster per degree-of-freedom (dof) than tensor-based tetrahedral (tet) operator
evaluation \cite{moxey20}.  By diagonalizing one direction at a time, the SEM
structure admits fast (exact or inexact) block solvers for local Poisson problems,
which serve as local smoothers for
$p$-multigrid (pMG) \cite{lottes05}.  $C^0$ continuity implies that the SEM is
{\em communication minimal:} data exchanges have unit-depth stencils,
independent of $N$.  Local $i$-$j$-$k$ indexing avoids much of the
indirect addressing associated with fully unstructured approaches, such that
high-order SEM implementations can realize significantly higher throughput
(millions of dofs per second) than their low-order counterparts
\cite{ceed_bp_paper_2020}. The $O(N)$ computational intensity of the SEM brings
direct benefits because its rapid convergence allows one to accurately simulate
flows with fewer gridpoints than lower-order discretizations.
Again in the regime of high-fidelity simulation of turbulence, the asymptotic error behavior of
the SEM, $C$=$O(h^N)$, manifests its advantage. It is more efficient to increase $N$
than to decrease the grid spacing, $h = O(E^{-\frac{1}{3}})$.

Codes that are comparable in scalability to NekRS include:
Nektar \cite{moxey20}, which employs the Nek5000 communication
           kernel, {\em gslib} and is based primarily on tets;
libParanumal\cite{ChalmersKarakusAustinSwirydowiczWarburton2020};
NUMO\cite{numo};
Neko\cite{neko};
deal.ii \cite{dealii}; and
MFEM \cite{mfem}, with the latter two having CPU kernel 
performance comparable to Nek5000 \cite{ceed_bp_paper_2020}.
% NUMO, which targets ocean modeling, is also OCCA based.
%     Neko is a Nek5000-based code under development at KTH in Stockholm.
%     libParanumal, which is also
%     under CEED, is at the cutting edge of GPU-based node performance
%     on NVIDIA and AMD platforms \cite{ceed_special_issue2}.

%%%%%%%%%%%%%%%%%%%%%%%%%%%%%%%%%%%%%%%%%%%%%%%%%%%%%%%%%%%%%%%%%%%%%%%%%%%%%

%%%%%%%%%%%%%%%%%%%%%%%%%%%%%%%%%%%%%%%%%%%%%%%%%%%%%%%%%%%%%%%%%%%%%%%%%%
%%\section{Innovations Realized}
\section{Advanced Algorithms}            
%%%%%%%%%%%%%%%%%%%%%%%%%%%%%%%%%%%%%%%%%%%%%%%%%%%%%%%%%%%%%%%%%%%%%%%%%%

A novel aspect of the current simulations is the use of {\bf high-order
overset grids on GPUs}, which has recently been ported to NekRS
following work by Lindquist \cite{lindquist21}.  The overset approach is
basically a Schwarz overlapping method applied to the NSE.
At each timestep, each {\em session} (or grid), $s$, advances the NSE
independently on domain $\Omega_s$ having boundary $\dO_s$.
Following \cite{mittal19b}, boundary data on segments of $\dO_s$
that intersect another domain, $\Omega_{s'}$, is computed as a $k$th-order
extrapolant (in time) of the data from $\Omega_{s'}$.  For stability, it is
necessary to iterate (i.e., via a predictor-corrector scheme) on the linear
Stokes subset of the NS advancement when $k>0$.  Simply using the data from the
prior timestep (i.e., $k=0$) is, however, sufficient for most turbulent flows
with no observable impact on turbulence statistics.  This approach was
developed in earlier work for Nek5000
\cite{peet12,merrill16,mittal19b,mittal20c}.  It is similar to the overset grid
technology that is well established in other codes (e.g.,
\cite{steger1983,bassetti98}) and that has been ported to GPUs in more recent
works (e.g.,
\cite{dmavriplis2013,witherden2020,overflow2024}).

\newcommand{\argminE}{\mathop{\mathrm{argmin}}}          % ASdeL

Central to the Nek5000/RS overset implementation is the scalable interpolation
library, {\em findpts()}, written by James Lottes.  Designed for millions of 
processors, this routine allows any MPI rank to post an interpolation query
for any $\bx^* \in \RR^d$ and returns to this rank the value of the requested
fields at $\bx^*$.  (Void if $\bx^* \not \in \Omega$.)
{\em findpts()} has three components: 
{\em (i)} a local setup, which identifies the mesh geometry on each rank; 
{\em (ii)} a nonlocal ``find-points'' call, which returns to the query-rank 
the MPI rank, element number, and local coordinates $\br^* \in \Oh$
corresponding to $\bx^*$;
and
{\em (iii)} an evaluation call, {\em findpts\_eval(u)}, which returns the value
of scalar or vector fields at $\bx^*$.  These calls are made {\em en masse}
(multiple queries from multiple processors) and
data exchanges are effected in $\log P$ time using fast implementations
of the generalized all-to-all {\em crystal router} routine of \cite{fox88}.
Global and local hash tables accelerate the search for the owning MPI rank
and owning element, respectively, that contains $\bx^*$. 
On candidate elements, $\Omega^e$, a Newton iteration identifies the minimizer,
$\br^*= \argminE_{\br \in \Oh} \| \bx^e(\br) - \bx^* \|$, to a small
multiple of machine precision such that successful interpolations yield
the full accuracy of the high-order expansion (\ref{eq:field1}).
For static domains, {\em (i)--(ii)} are called only once while {\em
findpts\_eval(u)} is invoked for each timestep.  

In our overset mesh setting, each session (subdomain) has a different MPI
communicator, save for the interpolation step, in which we use a global
communicator in order to leverage the scalable exchange utilities of {\em
findpts()}.  The principal work for {\em findpts\_eval()} is evaluation of
(\ref{eq:field1}) for each point on the owning processors.  Although the
interpolation cost is nominally $O(N^3)$ per point, the operation can be cast
as a {\tt dgemm} if there are multiple points per element.  In NekRS, the main
cost is packing and unpacking the relatively small sets of data associated with
the subdomain interface points.
  Sessions in the so-called NekNek coupling typically comprise multiple MPI ranks,
with a different communicator assigned to each session.  In effect, each session
is an independent fluid-thermal simulations.  The sessions can
differ in their polynomial orders, which means that each session runs an
independent executable (multiple-program, multiple-data parallelism).
They can also differ in the physical model for each.  For example, one 
session could run a RANS (Reynolds-Averaged Navier Stokes) model, while the
other runs LES.  

In the present case, we use one session to simulate a conjugate heat transfer
(CHT) problem consisting of coolant flow in the passage-ways, coupled with a thin
layer of surrounding elements that form part of the supporting solid walls.
Because the solid mesh for this simplified CHT problem is simply a skin
around the fluid mesh it is easy to construct.  To complete the problem,
the CHT domain is coupled via overset to a solid-only domain that includes both
steel and copper, which are illustrated by the grey and pink regions in Fig.
\ref{fig:chimera}.
{\em The overlapping, interpolation-based, coupling greatly simplifies meshing
of complex domains because the mesh topologies do not need to be conforming and
the surfaces do not need to match precisely.}
The solid session solves an unsteady conduction problem that is computationally
less demanding than the fluid-thermal CHT problem.  The sessions are synchronized
by data exchanges at the end of each timestep and the number of ranks assigned
to the solid session can be reduced to minimize wait (sync) time.
  For more general fluid-fluid couplings, details such as mass flux, volume
integrals, and multidomain ($> 2$) interpolation present additional technical isses
that have been addressed in \cite{mittal19b,mittal20c} that will be supported
in NekRS v24.

%%%%%%%%%%%%%%%%%%%%%%%%%%%%%%%%%%%%%%%%%%%%%%%%%%%%%%%%%%%%%%%%%%%%%%%%%%
\section{Performance Measurements}
%%%%%%%%%%%%%%%%%%%%%%%%%%%%%%%%%%%%%%%%%%%%%%%%%%%%%%%%%%%%%%%%%%%%%%%%%%

% Rod1717 with 170 layers         
\begin{figure*} \centering
      \includegraphics[width=3.1in]{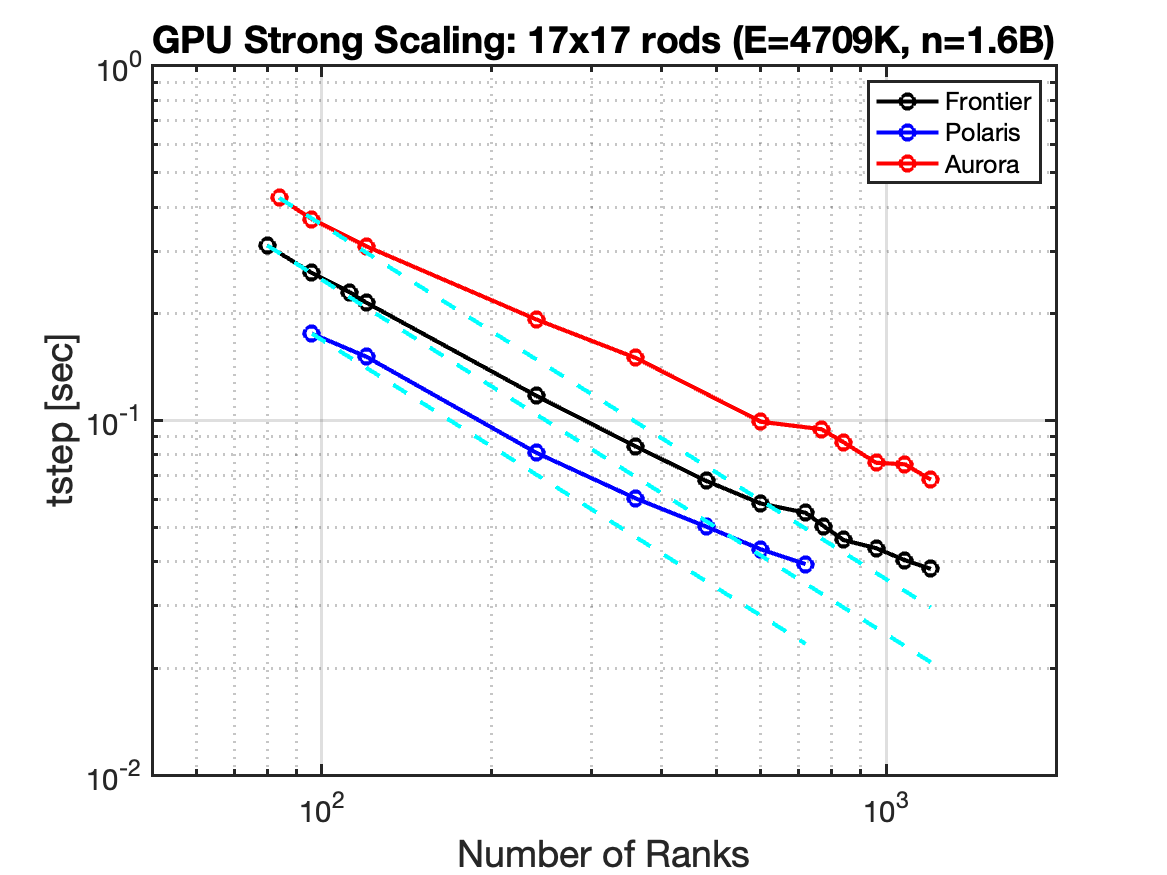}
      \includegraphics[width=3.1in]{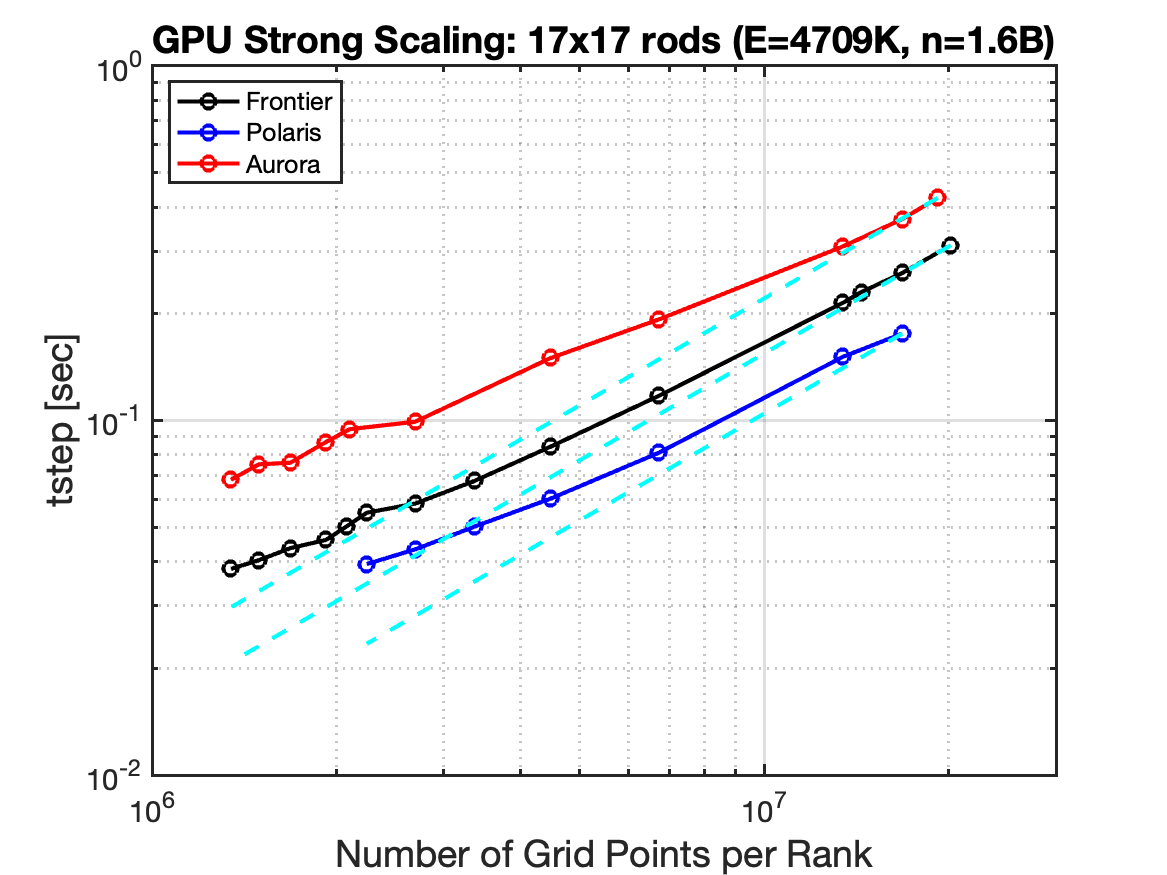}
      \\
      \includegraphics[width=3.1in]{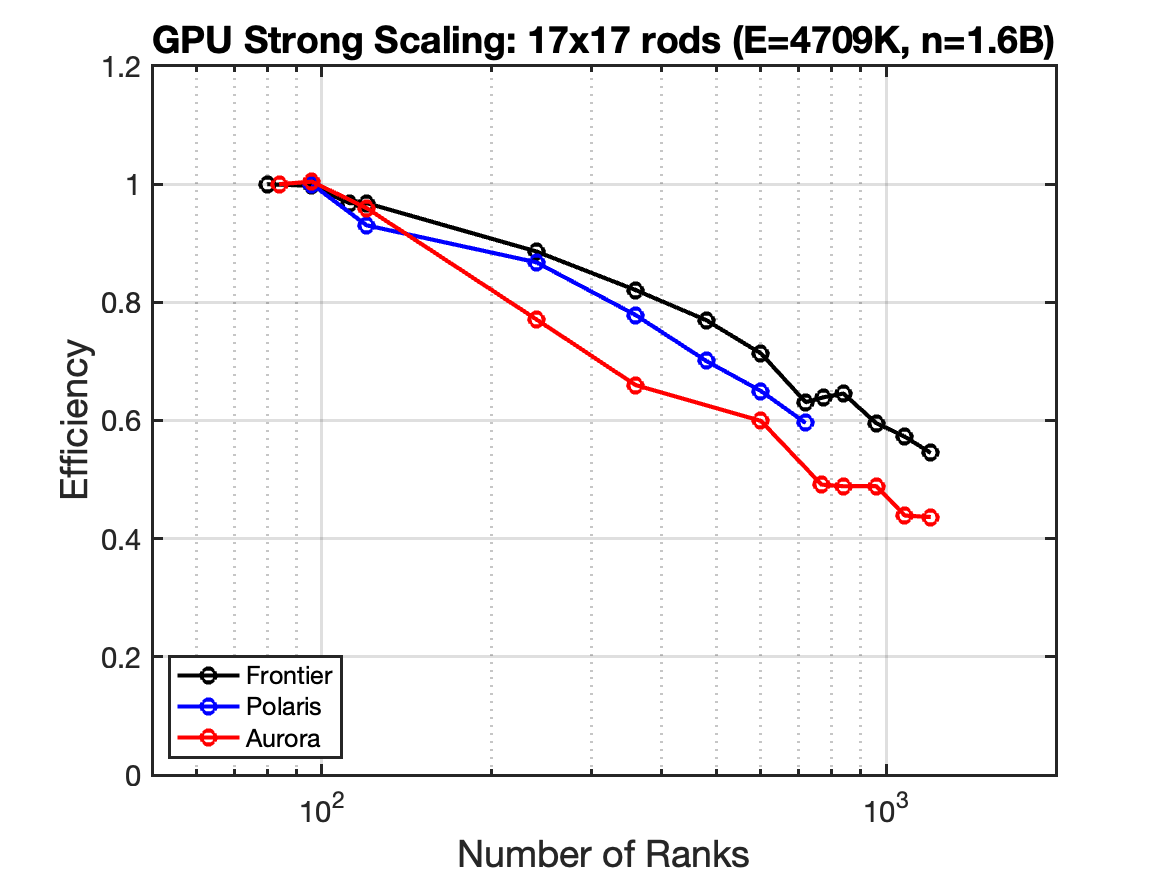}
      \includegraphics[width=3.1in]{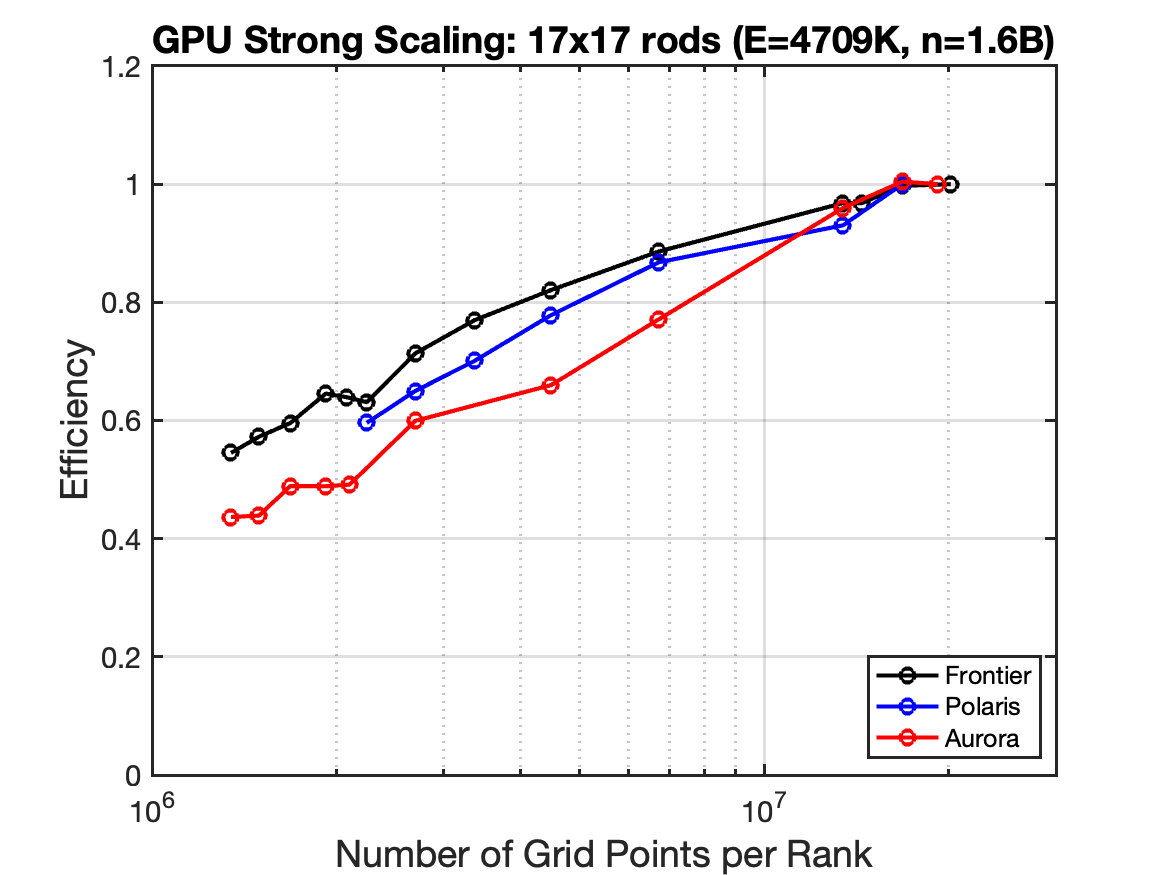}
      \\
      \includegraphics[width=3.1in]{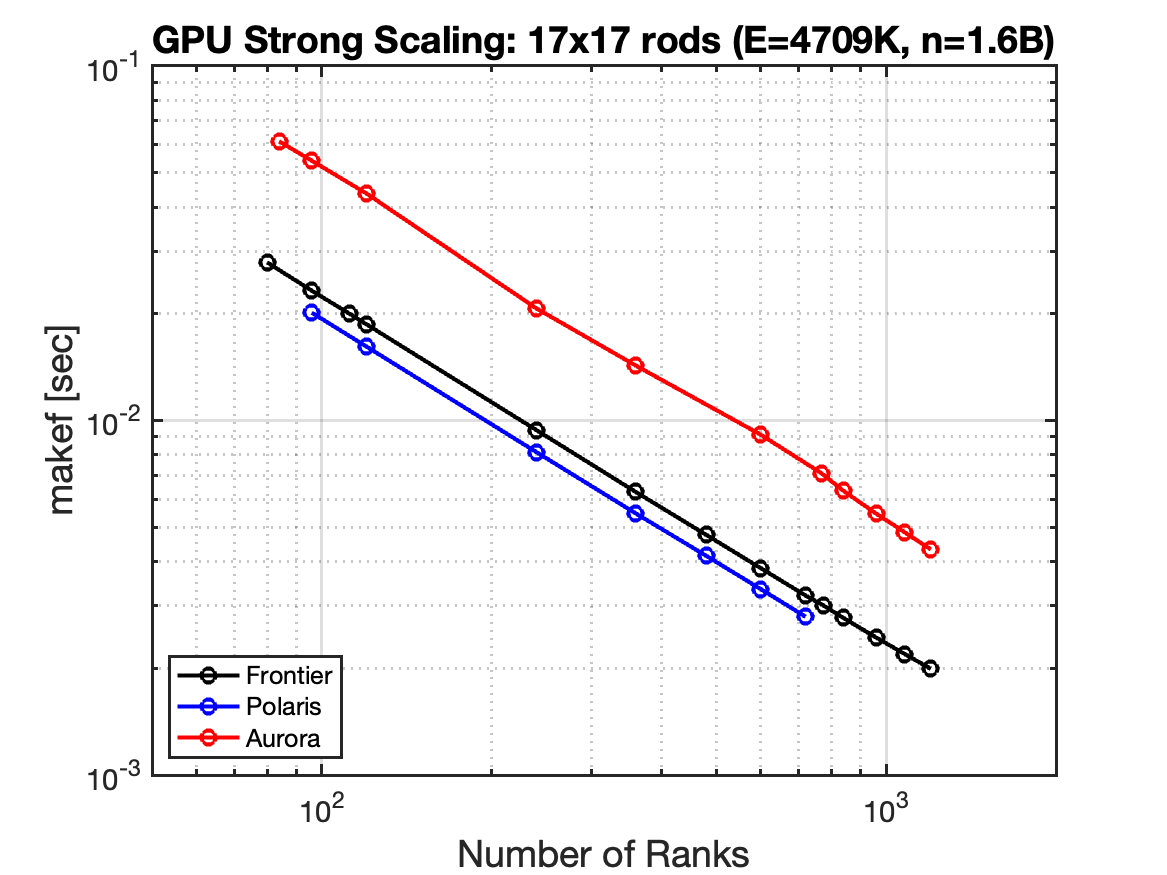}
      \includegraphics[width=3.1in]{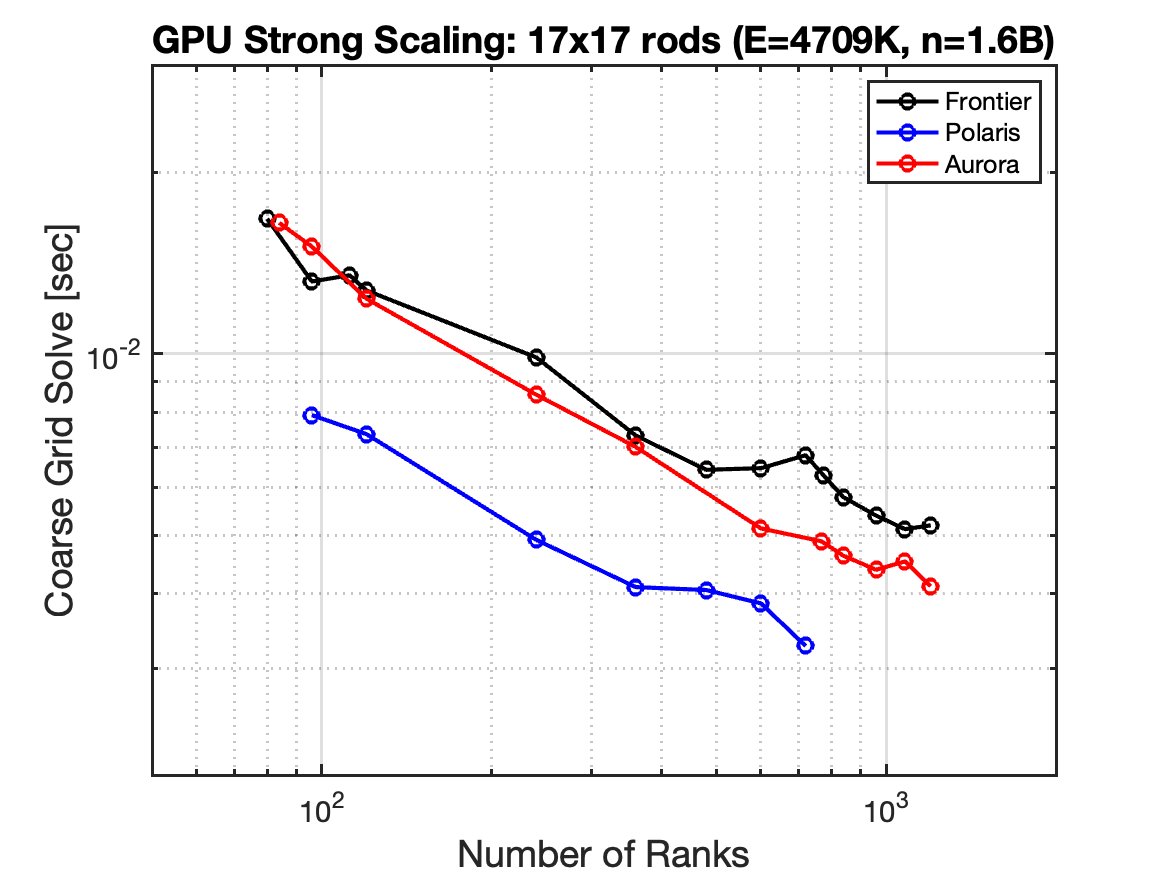}
% \caption{\label{fig:rod1717-170-scale} Aurora, Frontier, and Polaris strong-scaling 
  \caption{\label{fig:scale} Aurora, Frontier, and Polaris strong-scalings 
            for Navier-Stokes simulation of $17\times 17$ rod bundle using the total number of
            grid points, $n=1.6B$.
            The average (wall) time per step, tstep, in seconds is measured over steps 1001--2000.
            makef is a routine responsible for setting up the right-hand-sides for the momentum equations, 
            involving evaluation of the dealiased advection operator for the velocity.}
\end{figure*}

%%%%%%%%%%%%%%%%%%%%%%%%%%%%%%%%%%%%%%%%%%%%%%%%%%%%%%%%%%%%%%%%%%%%%%%%%
  \begin{table} %[!t]
    \caption{\label{tab:log1} NekRS statistics output for CHT (session 1)
case with 26,433,276 fluid elements and 6,440,484 solid elements, each
of order $N=5$, using $P=1096$ GCDs on Frontier.  The solid-only (session 2)
results used $E=55,056,380$ elements and $P=1720$ GCDs.
  }
  \scriptsize
\vspace*{-.2in}
  \begin{verbatim}
runtime statistics (step= 2000  totalElapsed= 1037.87s):

name                    time          abs%  rel%  calls
  solve                 3.99061e+02s  100.0
    min                 1.80199e-01s
    max                 6.63928e+00s
    flops/rank          3.56158e+11
    udfExecuteStep      2.82903e-01s   0.1        2000
    makef               5.84134e+01s  14.6        2000
    makeq               3.12773e+01s   7.8        2000
    udfProperties       1.85385e-02s   0.0        2001
    neknek              1.10285e+02s  27.6        2000
      sync              9.72651e+00s   2.4   8.8  2000
      exchange          9.96520e+01s  25.0  90.4  2000
        eval kernel     2.49230e+00s   0.6   2.5  4000
    velocitySolve       3.87565e+01s   9.7        2000
      rhs               5.77085e+00s   1.4  14.9  2000
    pressureSolve       1.09136e+02s  27.3        2000
      rhs               1.73350e+01s   4.3  15.9  2000
      preconditioner    7.43816e+01s  18.6  68.2  2461
        pMG smoother    3.40848e+01s   8.5  45.8  4922
        pMG smoother    1.49686e+01s   3.8  20.1  4922
        coarse grid     2.10594e+01s   5.3  28.3  2461
      initial guess     9.22916e+00s   2.3   8.5  2000
    scalarSolve         3.01516e+01s   7.6        2000
      rhs               4.96832e-01s   0.1   1.6  2000
  \end{verbatim}
  \end{table}
%%%%%%%%%%%%%%%%%%%%%%%%%%%%%%%%%%%%%%%%%%%%%%%%%%%%%%%%%%%%%%%%%%%%%%
  \begin{table} %[!t]
    \caption{\label{tab:log2} NekRS statistics output for solid (session 2)
using $E=55,056,380$ elements of order $N=5$ and $P=1720$ GCDs, corresponding
to session 1 of Table \ref{tab:log1}.
  }
  \scriptsize
\vspace*{-.2in}
  \begin{verbatim}
runtime statistics (step= 2000  totalElapsed= 1032.63s):

name                    time          abs%  rel%  calls
  solve                 3.98246e+02s  100.0      
    min                 1.71048e-01s
    max                 5.95790e+00s
    flops/rank          3.82066e+10
    udfExecuteStep      1.71539e-01s   0.0        2000
    makeq               1.90070e+00s   0.5        2000
      udfSEqnSource     9.52365e-02s   0.0   5.0  2000
    udfProperties       1.76741e-02s   0.0        2001
    neknek              1.85230e+02s  46.5        2000
      sync              9.12206e+01s  22.9  49.2  2000
      exchange          9.33105e+01s  23.4  50.4  2000
        eval kernel     3.43530e+00s   0.9   3.7  4000
    scalarSolve         3.76495e+01s   9.5        2000
      rhs               4.75687e-01s   0.1   1.3  2000
  \end{verbatim}
  \end{table}
%%%%%%%%%%%%%%%%%%%%%%%%%%%%%%%%%%%%%%%%%%%%%%%%%%%%%%%%%%%%%%%%%%%%%%
At the start of every simulation, NekRS identifies the fastest kernel 
for each operation through a combination of MPI\_Wtime and barriers.
In this way, OCCA kernels developed for new platforms do not supercede
existing ones unless they are actually faster for the given data/platform
configuration.  The test results are reported for principal kernels in each logfile
and include the operation, degrees-of-freedom per second (GDOFS), 
precision (FP32/64), bandwidth (GB/s), performance (GFLOPS), and chosen kernel
version.   Inspection of these tables is extremely useful in identifying
under-performing kernels as platforms change.

A similar battery of tests is used to select communication patterns.
For example, data might be packed and exchanged on the device using
GPU-aware MPI, or it might be packed on the device and exchanged via
the host, or packed and exchanged on the host.  (Short messages tend
to favor this latter option.)  The exchanges are tested stand-alone
or in conjunction with overlapped computation, depending on the kernel
in question.  Overlapping communication with computation typically 
yields a 10--15 \% reduction in Navier-Stokes solution time.
Similarly, using FP32 in the preconditioners reduces memory and
network bandwidth pressure and yields another 10--15\% savings.
(All reported FLOPS are for FP64, with FP32 floating point operations
counted as a 1/2 FLOP.)

NekRS is fully instrumented to track basic runtime statistics, which are output
every 500 time steps unless the user specifies otherwise.  Timing breakdowns
follow the physical substeps of advection, pressure, and viscous update, plus
tracking of known communication bottlenecks such as the coarse-grid solve for
the pressure Poisson problem.  For the overset grid case there are two (or
more) logfiles and the statistics report the synchronization (load imbalance or
wait-time) overhead as well as the exchange costs, which includes data packing
and interpolation.  
Table \ref{tab:log1} illustrates the output for an overset
grid case and how it can be used in guiding performance optimization.  
First off, we see that each of the 1096 GCDs (2 GCDs per AMD MI250X on
OLCF's Frontier supercomputer), sustains 356 GFLOPS, which is respectable,
particularly given the synchronization and communication overhead for the
overset grids.  In this case, we see that the nonlinear advection operator
({\tt makef}), which is
dealiased using $N_q=11$ quadrature points in each direction, consumes 14.6\%
of runtime, the overset grid exchange cost ({\tt neknek}) consumes 27.6\%, and
the pressure solve consumes 27.3\% of the time.  Here, there are no particular
outliers, save for the 27.6\% extra neknek overhead, which is the price paid
for having extreme geometric flexibility.  Note that, as demonstrated in
\cite{mittal19b}, the overhead on CPUs is generally lower ($< 10$\%).
Unstructured interpolation is not a GPU/SIMD-friendly operation.

Table \ref{tab:log2} shows the corresponding output for the solid-only domain,
which has $E=55M$.  The most striking feature of this case is that the
flop count is quite low, which indicates that fewer ranks could be effectively
assigned to this part of the domain.  

%%%%%%%%%%%%%%%%%%%%%%%%%%%%%%%%%%%%%%%%%%%%%%%%%%%%%%%%%%%%%%%%%%%%%%%%%%
\section{Performance Results}
%%%%%%%%%%%%%%%%%%%%%%%%%%%%%%%%%%%%%%%%%%%%%%%%%%%%%%%%%%%%%%%%%%%%%%%%%%

\noindent
{\bf Fission Results.}
From the logfile timing data, we plot strong-scaling performance in 
Fig. \ref{fig:scale} for a flow through a single $17 \times 17$ rod
bundle of a fission reactor for $E=4709$K elements of order $N=7$
($n$=1.6B gridpoints).
Shown in the top row
are measured values of the time-per-step for three platforms:
OLCF's Frontier, which features AMD MI250X GPUs, each having
two MPI ranks (one per GCD);
ALCF's NVIDIA A100-based Polaris (one rank per A100); and
ALCF's Aurora, which has Intel PVC GPUs, each running two MPI
ranks (one per tile).
We plot the performance as a function of 
$P$, the number of MPI ranks, in the upper left, and as
a function of $n/P$ on the upper right.
Because it factors out problem size, the number of points per rank is 
somewhat more universal than the number of ranks as an independent
variable \cite{ceed_bp_paper_2020}.
The plots show that Polaris is about $2\times$ faster than Aurora,
per rank and about $1.5 \times$ faster than Frontier.  We expect
that the Aurora numbers will improve as network issues are sorted out.
The center row shows the parallel efficiency. We see that, for this
case, Frontier realizes $\approx 80$\% efficiency at $n_{0.8}=n/P=4$M
points per rank, whereas Polaris has $n_{0.8} = 5$M, and Aurora
has $n_{0.8}=7$M.  Larger values of $n_{0.8}$ imply that one has
to use fewer processors, and thus run slower, to maintain 80\%
efficiency. (Small $n_{0.8}$ is better.)  In all likelihood, Aurora's
numbers will improve once this early system is debugged.

The last row of Fig. \ref{fig:scale} breaks out the costs of some of the
leading performance bottlenecks.  On the lower left, we plot the time-per-step
for the nonlinear advection evaluation, {\tt makef}.  We see that Aurora spends
roughly twice the amount of time in this kernel than either Frontier or
Polaris.  This kernel shows perfect linear scaling because it is compute
intensive and almost devoid of communication.  We speculate that the relatively
poor performance of Aurora stems from cache effects on the PVC.  Because
advection is dealiased with the 3/2's rule, the working data set per element is
$12^3$, rather than $8^3$, which is the size for the majority of the other
elemental operations.  Experiments in which we reduce the order of dealising
from 12 to 9 for this case led to a 1.5$\times$ reduction in advection time,
but this is not as dramatic an improvement as one might expect when cache
spilling is alleviated.  We note that the majority of the Aurora kernel
performance is on par with Frontier.

\begin{table*}
\centering
\caption{CHIMERA Cases: nelv=\# of fluid elements; nelt=\# of thermal-fluid elements}
\begin{tabular}{|c|c|c|c|c|c|c|c|}
\hline
\textbf{$E$} & \textbf{N} & \textbf{Type} & \textbf{dt} & \textbf{avg CFL} & \textbf{avg Pr} & \textbf{range} & \textbf{note} \\
\hline
  21M & 3 & overset & $1 \times 10^{-5}$ & 0.068 & 1 & 1001-2000 & \\ \hline
  21M & 4 & overset & $1 \times 10^{-5}$ & 0.112 & 1.046 & 1001-2000 & \\ \hline
  21M & 5 & overset & $1 \times 10^{-5}$ & 0.165 & 1.007 & 1001-2000 & \\ \hline
  21M & 5 & overset & $1 \times 10^{-4}$ & 1.483 & 1.046 & 1001-2000 & \\ \hline
  51M & 7 & fluid only & $1 \times 10^{-5}$ & 0.534 & 1 & 1001-2000 & \\ \hline
 211M & 7 & fluid only & $2 \times 10^{-4}$ & 1.087 & 5.216 & 1001-2000 & \\ \hline
%1.69B & 3 & fluid only & $1 \times 10^{-4}$ & 0.253 & 6.231 & 10001-15000 & \\ \hline
1.69B & 3 & fluid only & $5 \times 10^{-4}$ & 1.257 & 8.099 & 1001-2000 & pproj=8 \\ \hline
1.69B & 3 & fluid only & $5 \times 10^{-4}$ & 1.257 & 7.269 & 1001-2000 & pproj=L30 \\ \hline
1.69B & 7 & fluid only & $1 \times 10^{-4}$ & 1.103 & 50.195 & 101-300 & MG7531-cheb6 \\ \hline
1.69B & 9 & fluid only & $5 \times 10^{-5}$ & 0.872 & 120.55 & 21-60 & MG951-cheb3 \\ \hline
%1.69B & 9 & fluid only & $5 \times 10^{-5}$ & 0.872 & 68.725 & 21-60 & MG9751-cheb6 very slow \\ \hline
\end{tabular}
\label{tab:neknek2}
\end{table*}

On the lower right of Fig. \ref{fig:scale} we plot the coarse-grid solve
time per step, which can be significant for large values of $P$ \cite{sc22}.
In contrast to advection, the coarse-grid solve is communication intensive
and uses Hypre on the host CPU to solve the unstructured sparse $p=1$ problem,
which has $\approx E$ unknowns.  Here, Polaris is $\approx 2\times$ faster
than either Aurora or Frontier.  Remarkably, this result indicates that
Aurora's host-to-host network is reasonably fast and perhaps running better
than that of Frontier.
\\

\noindent
{\bf Fusion Results.}
We have constructed a sequence of CHIMERA meshes with and without the overset
solid geometry.  A subset of these is listed in Table \ref{tab:neknek2}.  We
remark that the smallest of these cases, with $E=21$M elements is almost 50\%
larger than the record Nek5000 problem size  of $E=15$M, which ran on 1M ranks
of ALCF's BG/Q, Mira, at the start of the ECP project in 2017.  The sequence
in Table \ref{tab:neknek2} is remarkable simply for the shear size, which 
scales up to $>$ 1 {\em trillion} grid points for the largest case.
These meshes were constructed using a tet-to-hex decomposition and the
largest two cases were generated using oct-refinement of the next smallest ones.

The table indicates that the number of pressure iterations per step is
reasonable ($< 10$) up to $E=1.69$B, $N=7$ ($n=580$B), where it climbs to $>
50$ and ultimately to $> 120$ iterations per step.  Unfortunately, these larger
cases are full Frontier runs and thus challenging to debug/optimize.
We had some success, however, for the $E=1.69$B, $N=3$ ($n=45$B) case, which
could be run on a reasonably small subset of Frontier.  At each timestep
in Nek5000/RS, we generate an initial guess for the pressure (and velocity)
by projecting onto the space of prior solutions \cite{fisc98}.  This approximation
is stable and exponentially convergent with the number of prior solutions saved,
with accuracy ultimately controlled by the iteration tolerance.  By increasing
the number of prior solutions from 8 to 30, the average pressure iteration
was reduced from 8.1 to 7.3 to yield an overall 10\% reduction in simulation time.
There is some hope that similar gains could be realized for the larger cases
in the table, particularly where the pressure iteration counts are large.
Also, using aggressive $p$-multigrid schedules of $p=7$, 5, 3, and 1 with
6th-order Chebyshev smoothing made inroads for the $E=1.69$B, $N=7$ case.
Nonetheless, the 1 to 2 orders-of-magnitude increase in pressure iterations
is a signifcant barrier to overcome for problems at this scale; just discovering
whether the issue lies in a bug in the mesh, the solver, or the algorithm
will require significant human and computational resources.
Finally, we remark that the performance for the fluid only cases 
ranges from 160 GFLOPS/rank for the $E=$1.6B cases on $P=73728$ ranks
of Frontier (12 PFLOPS total) to 700 GFLOPS/rank for the $E=$51M
case on 4000 ranks (2.8 PFLOPS total).  The fall-off in per-rank
performance is attributable in part to network noise on Frontier
and is alleviated to some extent by bypassing the GPU-direct option
in NekRS.

\begin{figure*} \centering
      \includegraphics[width=2.93in]{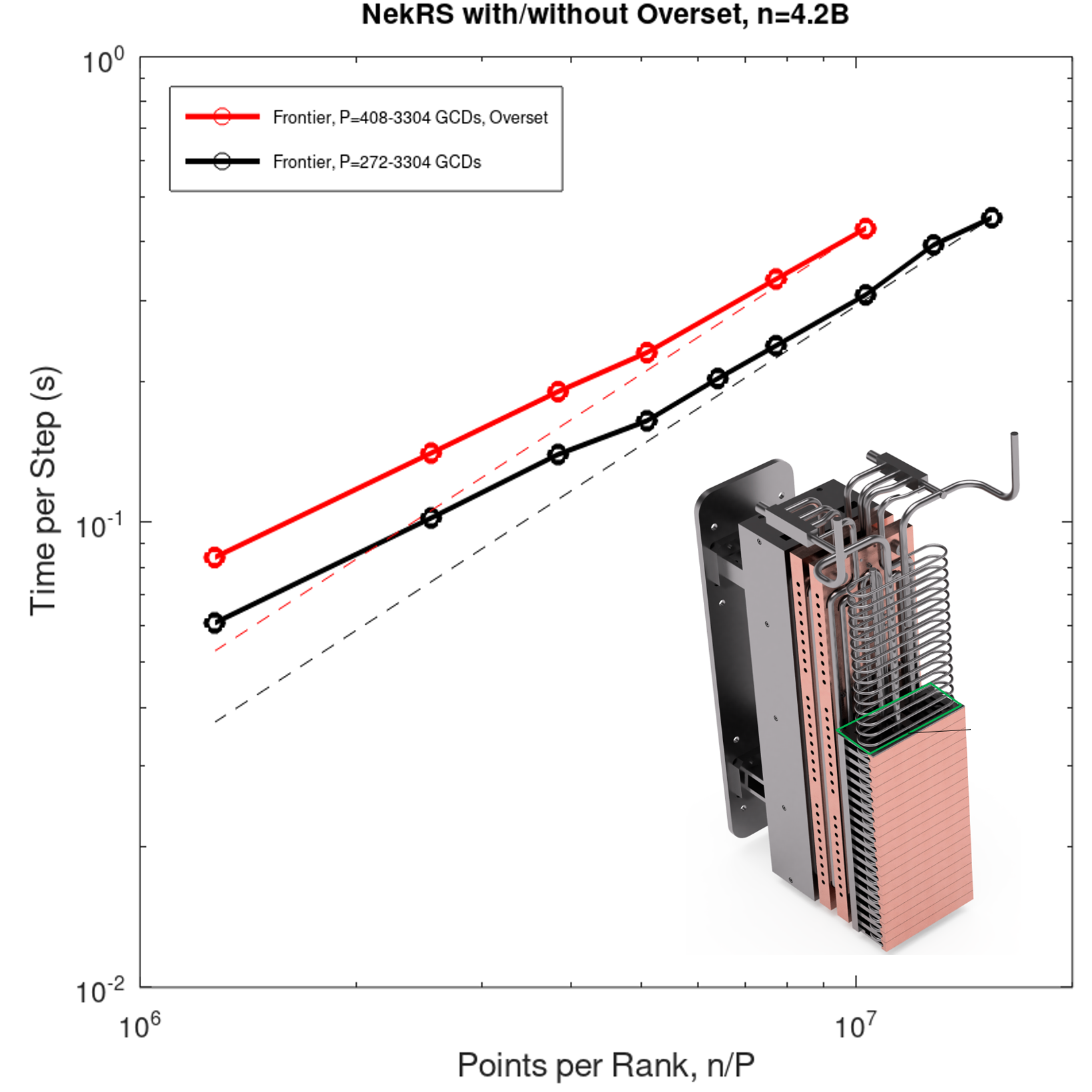}
\hspace{.33in}
      \includegraphics[width=2.8in]{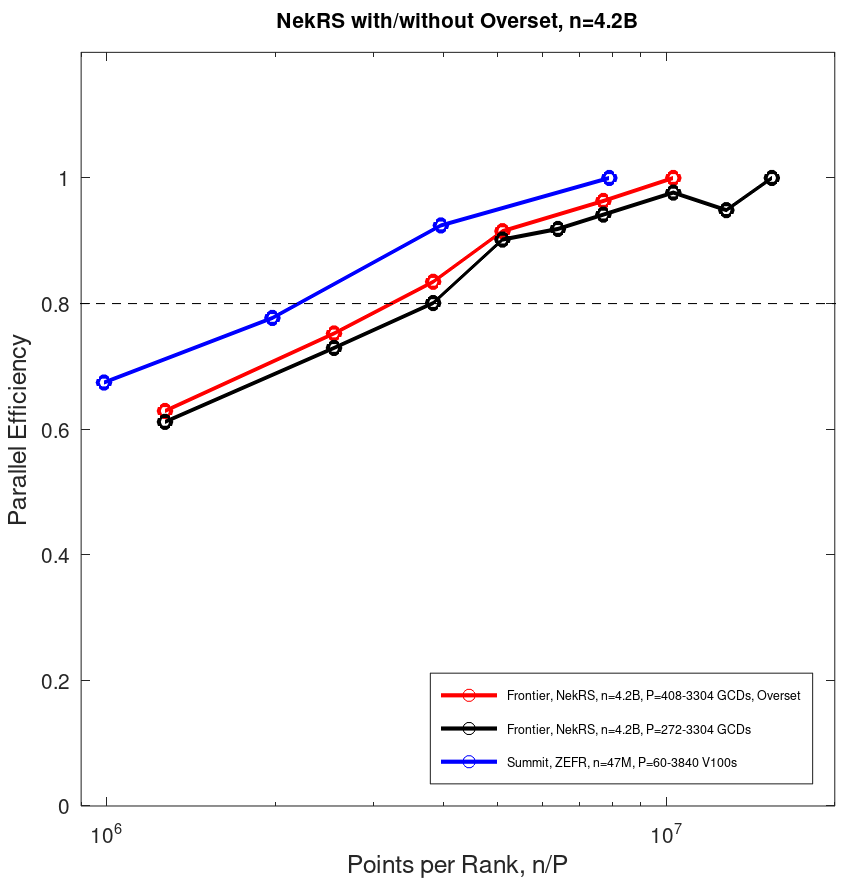}
  \caption{\label{fig:neknek} 
    Overset grid results for the CHIMERA geometry (inset).
Left: strong-scaling time-per-step for the fluid-thermal (CHT)
part of the geometry with and without overset coupling.
Right: parallel efficiency for the same cases; also shown
is the efficiency for the comparable ZEFR code ($N=4$)
running on the V100s of Summit \cite{witherden2020}.
   }
\end{figure*}

Next, we examine strong scalability for overset grids on GPUs for the
CHIMERA geometry.
Figure \ref{fig:neknek} (left) shows the time-per-step for a case with 
$E=26,433,276$ fluid elements and 6,440,484 solid element in session 1.
Session 2 has 55,056,380 solid elements.  Also shown is a simple
monodomain CHT case that is the same as session 1 of the overset
grid case.  We see that the monodomain case requires about 25\%
less time per step than the overset grid case.
Figure \ref{fig:neknek} (right) shows the parallel efficiency
for these two cases, which indicates that one can use roughly
the same number of processors for the overset case as for the
monodomain case at the 80\% efficiency mark.
Also shown is the efficiency for the comparable ZEFR code ($N=4$)
running on the V100s of Summit \cite{witherden2020}.

% \begin{table*}%[t]
% \centering
% \caption{CHIMERA Cases: nelv=\# of fluid elements; nelt=\# of thermal-fluid elements}
% \begin{tabular}{|c|l|r|c|c|} \hline
% \textbf{Case} & \textbf{Type} & \textbf{nelv (cht)} & \textbf{nelt (cht)} & \textbf{nelt (solid)} \\ \hline
% 13M & fluid only & 13,579,540 & =nelv & --- \\ \hline
% 26M & overset& 26,433,276 & 32,873,760 & 55,056,380 \\ \hline
% 51M & overset & 51,011,162 & 61,946,804 & 55,056,380 \\ \hline
% 211M & fluid only & 211,466,208 & =nelv & --- \\ \hline
% 329M & fluid only & 329,135,376 & =nelv & --- \\ \hline
% 1.69B & fluid only & 1,691,729,664 & =nelv & --- \\ \hline
% \end{tabular}
% \label{tab:neknek1}
% \end{table*}

%%%%%%%%%%%%%%%%%%%%%%%%%%%%%%%%%%%%%%%%%%%%%%%%%%%%%%%%%%%%%%%%%%%%%%%%%%
\section{Implications}
%%%%%%%%%%%%%%%%%%%%%%%%%%%%%%%%%%%%%%%%%%%%%%%%%%%%%%%%%%%%%%%%%%%%%%%%%%

\noindent
{\bf Fusion and Fission system designs.}
The simulations of reactor cores and fusion system devices described here are ushering in a
new era for the thermal fluid analysis of such systems. The possibility of simulating these systems in all their size
and complexity at this level of fidelity was unthinkable until recently. The simulations are already being used to benchmark and improve predictions obtained with traditional methods.  This innovation is particularly crucial for fusion blanket systems due to their complexity, lack of existing experimental datasets, and prohibitive heat transfer conditions. Modeling and simulation tools, once validated, can provide a powerful tool to complement experimental data and enable a broader exploration of the design space. We envision that the impact will be increasingly felt as fusion transitions from a physics challenge problem to an engineering challenge problem. This will, in turn, broadly serve the goal of reaching a carbon-free economy within the next few decades.
\\

\noindent
{\bf HPC, Algorithms, and CFD.}
This study demonstrates the continued importance of numerical algorithms and
implementations for HPC, with a several-fold increase in modeling size over a
short time frame.  Simulations have also increased in complexity (intricate
geometry, large problem sizes, overset grids). \textit{Attention to maintaining
a performant code base while increasing code and simulation complexity has
proven to be essential.}  In particular, Automated tuning has proven to be a
boon to performance portability since no single approach works for every
problem in the parameter space (e.g., spanning polynomial order, local problem
size, hardware and software version) as complexity increases.  For users, who
often have a singular interest, being able to deliver best-in-class performance
can be critical to productivity.   In Nek5000 and NekRS, we support automated
tuning of kernels and communication strategies that adapt to the architecture,
problem layout, network, and underlying topology of the particular graph that
is invoked at runtime.  Performance gains realized through this study will be
leveraged by the 500+ members of the Nek5000/RS user community and will, we
hope, establish an approach to portable HPC software running on everything from
workstations to exascale platforms and beyond.

%%- - - - - - - - - - - - - - - - - - - - - - - - - - - - - - - - - - - - - -%%

%%- - - - - - - - - - - - - - - - - - - - - - - - - - - - - - - - - - - - - -%%
%%%%%%%%%%%%%%%%%%%%%%%%%%%%%%%%%%%%%%%%%%%%%%%%%%%%%%%%%%%%%%%%%%%%%%%%%%%
\section*{Acknowledgments}
%%%%%%%%%%%%%%%%%%%%%%%%%%%%%%%%%%%%%%%%%%%%%%%%%%%%%%%%%%%%%%%%%%%%%%%%%%%
    This material is based upon work supported by the U.S. Department of Energy,
    Office of Science, under contract DE-AC02-06CH11357  and by the Exascale
    Computing Project (17-SC-20-SC), a collaborative effort of two U.S.
    Department of Energy organizations (Office of Science and the National
    Nuclear Security Administration). The research used resources at the
    Argonne Leadership Computing Facility at Argonne National Laboratory
    which is supported by the Office of Science of the U.S. Department of Energy
    under Contract DE-AC02-06CH11357 
    and the Oak Ridge Leadership Computing Facility at Oak Ridge National Laboratory,
    which is supported by the Office of Science of the U.S. Department of Energy
    under Contract DE-AC05-00OR22725.

%%- - - - - - - - - - - - - - - - - - - - - - - - - - - - - - - - - - - - - -%%

%%- - - - - - - - - - - - - - - - - - - - - - - - - - - - - - - - - - - - - -%%
%\bibliographystyle{elsarticle-num-names}
 \bibliography{bibs/emmd,bibs/tim}
%%- - - - - - - - - - - - - - - - - - - - - - - - - - - - - - - - - - - - - -%%

%%---------------------------------------------------------------------------%%
%%---------------------------------------------------------------------------%%
\end{document}